# NON-CLASSICALITY IN MENTAL STATES: AN EXPERIMENTAL STUDY WITH AMBIGUOUS AUDIO (MUSIC) STIMULI

Souparno Roy*, Ranjan Sengupta, Tarit Guhathakurata and Dipak Ghosh

Sir C. V. Raman Centre for Physics and Music, Jadavpur University, Kolkata – 700032
*thesouparnoroy@gmail.com

## ABSTRACT
*This paper attempts to address the question that whether the present physical or mathematical theories are sufficient to understand the complexities of human brain when it interacts with the external environment in the form of an auditory stimulus. There have been efforts reporting that the introduction of ambiguity in visual stimuli causes effects which can't be explained classically. In this paper, it is investigated whether ambiguity in auditory stimuli can introduce any non-classical effects in human brain. Simple experiments were performed on normal subjects where they listened to an ambiguous auditory signal and responded to a question with 'yes' or 'no'. The outcome of the test showed that the classical formula of total probability does not hold true in this case. Results were interesting and indicate that there is a definite non-classicality in mental states in perception of ambiguous audio stimuli.*

## INTRODUCTION
The brain is a complex, self-organized state of biological material [1]. How it works or interacts with external environment is still not well understood [2]. For a general idea, it is known that sensory organs in human body receive external experiences in various forms of stimuli (visual, auditory, olfactory, tactile), which then gets transported to the brain through neurons via electrical impulses. But detailed researches on how and why 'perception', 'cognition' or 'consciousness' gets generated by these nerve impulses are still in a nascent stage. There have been a number of theories throughout the years about the working principle of the most complex biological organ in humans. For some, it is a complex neural network obeying classical physics [3] [4] [5]. On the other hand, some authors argued that non-classical effects (such as quantum physics, macroquantum effects like superconductivity, Bose-Einstein condensate, superflourescence or some other mechanisms) are responsible for the brain to work [6] [7] [8] [9] [10] [11] [12] [13] [14]. It is evident that to know the intricate details of complex neural pathways, the interaction between brain and the environment is need to be examined thoroughly. In this study, the work is focused on exploring the effects of an ambiguous auditory stimulus in human perception. Auditory stimulus, especially noise free music, is a great agent in highlighting cognitive phenomenon. It is found that the brain's ability to absorb and make sense of music, what sometimes is referred to as organized sound, is highly complex and far more effective than even a computer's ability to process so [15] [16].

In cognitive research, a fundamental assumption is that it is possible to model cognition on the basis of formal, probabilistic principles. And the human cognitive models have long been explained using classical Bayesian probability theory. [17] [18] [19] [20]. But, a large body of work also reported findings which can't be explained using classical regime [21] [22] [23]. Many of these relate to order/context effects, violations of the law of total probability (which is fundamental to Bayesian modeling), and failures of compositionality. In recent times, a section of physicists and mathematicians have started working on the how the concept of classical probability is violated when subjects are exposed to ambiguous visual stimuli [24] [25]. They have found evidence of this violation and suggested that an extra term (claimed as 'interference term') is to be added to counter the effect [26]. Hence, in light of these recent reports, cognitive models of how music (or external information, in general) affects brain need to be reviewed. The aim of this work is to investigate whether human brain shows any such effects, non-explainable classically, when it is stimulated with an external ambiguous music.

## THEORETICAL BASIS OF THE EXPERIMENT

To agree with this study, one must remember the point that, to examine non-classicality, violation of Bayes' formula for total probability is the first stepping stone. Bayesian probability holds true in an exhaustive sample space where events are mutually exclusive. Combining events obey to logic of set theory (conjunction and disjunction operations) and to the distributive axiom. Now, the formula for total probability is:

**p (A±) = p (B+).p (A±/B+) + p (B-).p (A±/B-)** ; where : (i) A and B are two dichotomous random variables having values A+,A- and B+,B- respectively , (ii) p(x) is the probability of event x and (iii) p(x/y) is the conditional probability of event x assuming event y happened already.

This is a formula having sum of discrete probability values of discrete events. Classically, it means that if two variables are measured with their respectively weighed percentages, no matter what their order or orientations are, the desired result will always match the above formula. In the previous work with ambiguous visual stimuli [25], the apparent ambiguity invokes two responses for a specific question (For example, in experiment 1 of [25], the answer to the question 'whether the two lines have same lengths' are 'yes' and 'no'). These responses serve as variables. Now the probabilities of occurring these two results are calculated in different orientations. The results are shown to violate classical Bayesian theory, according to which the results should always follow the formula in all the orientations of the experiment.

Here it is sought to examine whether an induced dichotomy in auditory signal makes the cognitive properties adhere to this classical notion of total probability. The subjects are divided in two ensembles and are exposed to two experimental conditions. To one group, task A is performed, response of which can either be 'yes' or 'no'. As a result, a probability p(A=+1) for 'yes' and a probability p(A=-1) for 'no' is obtained. To the second group, task B is performed first and immediately after that, task A follows. Here also, responses given by the subjects to each task are recorded as either 'yes' or 'no'. This gives probabilities p(B=+1), p(B=-1), p(A=+1/B=+1), p(A=-1/B=+1), p(A=-1/B=+1), p(A=-1/B=-1). The responses given have no ambiguities in themselves, only two discrete probabilities of discrete events 'yes' and 'no' can be obtained. Hence, if classical theories hold true, the collected data should match the law of total probability. This is the theoretical base upon which the experiment has been designed.

## EXPERIMENTAL DETAILS

The auditory stimuli used in this consisted of two music clips of 30 seconds stitched together with a silent rest period of 5 seconds in between. These two stitched clips have the same value of tempo, i.e., beats per minute or BPM. Here, the choice of the stimuli and the nature of the asked question should be explained. In previous reports on perceptual ambiguity, all the stimuli used are mainly visual [25] [27]. Auditory ambiguity is evidently hard to create because unlike visual stimulus, all audible signals have temporal components instead of spatial ones. Also, listening to two simultaneous clips can lead to incomprehensible noise. Considering these factors, the stimuli used are successive music clips which are otherwise dissimilar except one musical element (beats per minute). It created an ambiguity for the listener that whether the clips are different altogether or may be similar in one or more aspects. The resting period of 5 seconds is sufficiently small so that while listening to the latter clip, the effect of the former doesn't fade away. Thus, the ambiguity introduced is temporal by nature. Also, from the well known reports in music cognition it is established that for the human brain, the fastest and most distinguishable arousal occurs due to fluctuations in tempo than melody or rhythm or any other basic elements of music [28] [29] [30] [31] [32] [33]. Hence, the question posed was decided to be based on tempo. To eliminate musician/non-musician bias, subjects were asked not to use any means of tempo measurements, including finger/foot-tapping.

## METHOD OF ANALYSIS

In the study, 32 subjects were taken and are divided in two equal groups (say Group 1 and Group 2). Group 1 was exposed to a music clip as auditory stimuli (clip A) for 65 seconds. After that, subjects were asked whether the two clips they heard had the same tempo or not. The answer 'yes' to the posed question on clip A gave p(A=+1), and 'no', gave p(A=-1). The group 2 was made to listen to another similarly made clip (clip B) first, and immediately afterwards they were exposed to clip A (after 60-80 seconds). Responses of this group gave the remaining sets of probabilities, i.e., p(B=+1), p(B=-1), p(A=+1/B=+1), p(A=-1/B=+1), p(A=-1/B=+1), p(A=-1/B=-1) . The participants belonged to various age groups and of various music tastes. All had normal hearing capabilities (headphones were used). Finally, in all the cases, to avoid the risk to influence the subjects, the question to be asked by tests was posed in the most neutral form. With this data, we estimated the validation of the formula of total probability in case of auditory ambiguity.

## RESULTS AND DISCUSSION

The results are given below in Table 1. Table 2 gives the values of calculated probabilities.

**TABLE 1:** Results obtained for tests A, B and A/B

| Group 1 subjects | A+ | A- | | | Group 2 subjects | B+ | B- | | A+ | A- |
|---|---|---|---|---|---|---|---|---|---|---|
| 1 | | * | | | 1 | | * | | | * |
| 2 | | * | | | 2 | | * | | * | |
| 3 | | * | | | 3 | | * | | * | |
| 4 | * | | | | 4 | | * | | | * |
| 5 | | * | | | 5 | | * | | * | |
| 6 | * | | | | 6 | * | | | | * |
| 7 | * | | | | 7 | | * | | | * |
| 8 | | * | | | 8 | | * | | | * |
| 9 | | * | | | 9 | * | | | * | |
| 10 | * | | | | 10 | | * | | | * |
| 11 | * | | | | 11 | | * | | | * |
| 12 | * | | | | 12 | | * | | | * |
| 13 | | * | | | 13 | | * | | * | |
| 14 | * | | | | 14 | * | | | | * |
| 15 | | * | | | 15 | | * | | | * |
| 16 | | * | | | 16 | | * | | | * |

**TABLE 2:** Obtained values of probabilities

| | |
|---|---|
| P(A+) | 0.4375 |
| P(A-) | 0.5625 |
| P(B+) | 0.1875 |
| P(B-) | 0.8125 |
| P(A+/B+) | 0.3333 |
| P(A+/B-) | 0.3077 |
| P(A-/B+) | 0.6667 |
| P(A-/B-) | 0.6923 |
| P(A+)=P(B+)*P(A+/B+) + P(B-)*P(A+/B-) | 0.3125 |

| P(A-)=P(B+)*P(A-/B+) + P(B-)*P(A-/B-) | 0.6875 |

From the collected data, we found that in Group 1, the probability of A+ is **0.4375** and in case of A- it is **0.5625**.

In group 2, the probability data are: B+ is **0.1875**, B- is 0.8125, p(A+/B+) is **0.3333**, p(A+/B-) is **0.3077**, p(A-/B+)= **0.6667** and P(A-/B-) = **0.6923**.

Now, using the law of total probability, i.e., $p(A\pm) = p(B+).p(A\pm/B+) + p(B-).p(A\pm/B-)$, the calculated probability of A+ is found to be **0.3125** and for A- it is **0.6875**.

Fig. 1 shows the difference in probability values of A+ and A- for the two groups.

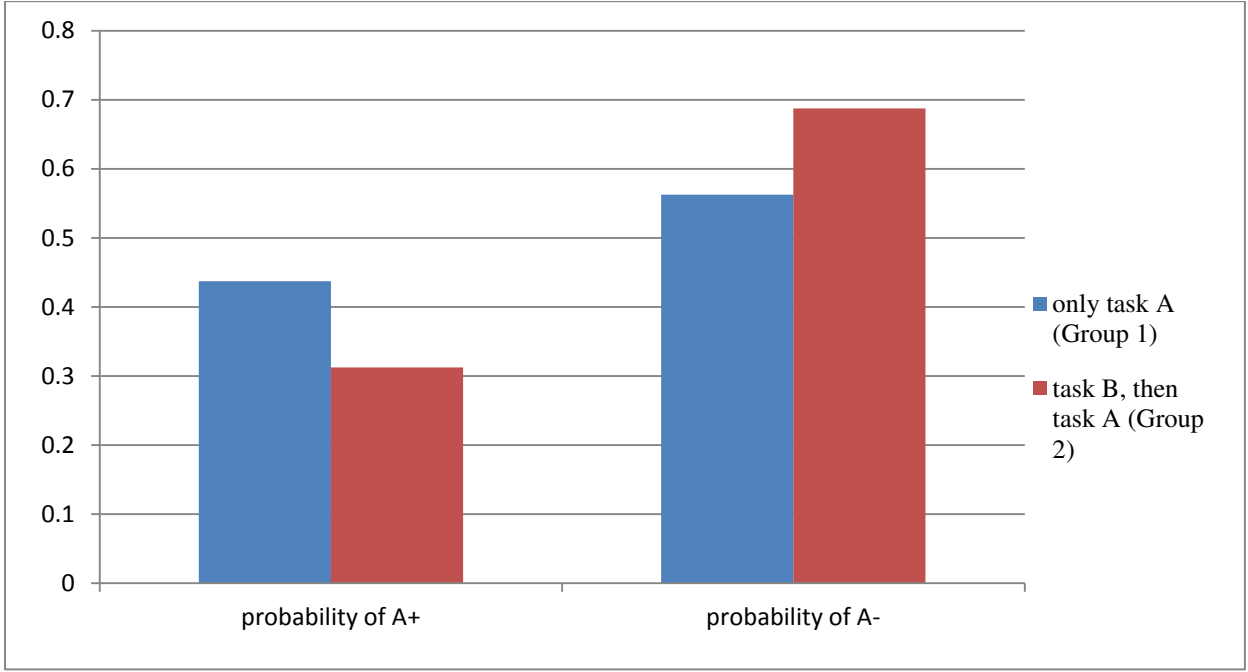

**Figure 1 :** Difference in values of calculated probabilities

The difference in the calculated value of Group 1 and Group 2 is: $|\Delta p (A\pm)| =$ **0.1250.**

The data shows clearly there is a difference between values obtained from two experimental calculations. Thus, the formula of total probability is seem to be violated and an extra term of $|\Delta p (A\pm)|$ has crept up which is completely unaccounted for in classical theoretical regime.

## CONCLUSION

The present study yields a result which indicates towards the violation of classical probability laws in case of ambiguous auditory signals. This observation is in agreement with previous reports using visual stimuli. The resulting inconsistency between the probability data of two different group tasks may lead to the agreement on a modified nondeterministic approach over the classical probability model as it is based on the sum of discrete single event probabilities and doesn't include contextuality, interference or fuzziness. In addition, the significant difference in the values of the probability function in the second experiment over first one might be due to some contextual effect introduced by the second clip; this needs to be explored in detail. Finally, we can conclude that this study provides significant information provoking further investigations on the nature of the non-classicality in cognitive properties.